\documentclass[conference, a4paper]{IEEEtran}

\usepackage{cite}
\usepackage{amsmath,amssymb,amsfonts}
\usepackage{graphicx}
\usepackage{textcomp}
\usepackage{xcolor}
\usepackage{listings}
\usepackage{booktabs}
\usepackage{multirow}
\usepackage{tikz}
\usetikzlibrary{arrows.meta,positioning,shapes.geometric,fit,calc,backgrounds}
\usepackage{pgfplots}
\pgfplotsset{compat=1.18}
\usepackage[hidelinks]{hyperref}
\usepackage{balance}

\begin{document}

\title{SHADOW: Seamless Handoff And Zero-Downtime\\Orchestrated Workload Migration\\for Stateful Microservices}

\author{
\IEEEauthorblockN{Hai Dinh-Tuan}
\IEEEauthorblockA{
\textit{Technische Universit\"at Berlin} \\
Berlin, Germany \\
hai.dinh-tuan@tu-berlin.de}
}

\maketitle

%
\begin{abstract}
Migrating stateful microservices in Kubernetes requires careful state
management because in-memory state is lost when a container restarts.
For StatefulSet-managed workloads, the problem is amplified by identity
constraints that prohibit two pods with the same ordinal from running
simultaneously, forcing a sequential stop-restore cycle with unavoidable downtime. This paper presents SHADOW
(\textbf{S}eamless \textbf{H}andoff \textbf{A}nd Zero-\textbf{D}owntime
\textbf{O}rchestrated \textbf{W}orkload Migration), a Kubernetes-native
framework that implements the Message-based Stateful Microservice Migration
(MS2M) approach as a Kubernetes Operator. SHADOW introduces the
\textit{ShadowPod} strategy, where a shadow pod is created from a CRIU
checkpoint image on the target node while the source pod continues serving
traffic, allowing concurrent operation during message replay. For
StatefulSet workloads, an identity swap procedure with the ExchangeFence
mechanism re-checkpoints the shadow pod, creates a StatefulSet-owned
replacement, and drains both message queues to guarantee zero message loss
during the handoff. An evaluation on a bare-metal Kubernetes cluster with
280 migration runs across four configurations and seven message rates shows that, compared to the sequential baseline on the same
StatefulSet workload, the ShadowPod strategy reduces the restore phase by up
to 92\%, eliminates service downtime, and reduces total migration
time by up to 77\%, with zero message loss across all 280 runs.
\end{abstract}


\section{Introduction}
\label{sec:introduction}

The increasing adoption of microservice architectures in cloud-native and
compute continuum environments \cite{11410125} has introduced new challenges around service
mobility. While stateless services can be rescheduled across cluster
nodes, stateful microservices that maintain in-memory state for
performance-critical operations require careful state management during
migration. Traditional migration techniques, originally developed for virtual
machines, treat service state as an opaque memory block and rely on pre-copy or
post-copy mechanisms that introduce additional complexity and network
overhead~\cite{ma2019live, guitart2024postcopy}.

The Message-based Stateful Microservice Migration (MS2M)
framework~\cite{dinhtuan2022ms2m} addresses this challenge by using the
application's own messaging infrastructure to reconstruct state on the target
node. Rather than copying raw memory, MS2M checkpoints the running container,
restores it on a target node, and replays buffered messages to synchronize
state, reducing downtime to the checkpoint phase. A subsequent
extension~\cite{dinhtuan2025k8s} integrated MS2M into Kubernetes using Forensic
Container Checkpointing (FCC)~\cite{k8sfcc}, introduced a threshold-based cutoff
mechanism for bounded replay times, and adapted the procedure for StatefulSet
workloads using a Sequential (stop-then-restore) strategy.

However, the Kubernetes integration in~\cite{dinhtuan2025k8s} exhibited two
performance bottlenecks. First, the restore phase for StatefulSet
pods required a median of 38.5 seconds due to Kubernetes' identity constraints,
which prohibit two pods with the same StatefulSet-managed hostname from running
simultaneously. This forced a sequential stop-restore cycle, during which the
service was unavailable. Second, the migration was orchestrated
by an out-of-band agent (\texttt{Migration Manager}) that communicated with the cluster via \texttt{kubectl}, operating outside Kubernetes' declarative resource model.

This paper presents SHADOW (Seamless Handoff
And Zero-Downtime Orchestrated Workload Migration), a Kubernetes-native
framework that addresses these limitations through three contributions:

\begin{enumerate}
\item \textit{Kubernetes Operator:} a Custom Resource Definition
  (\texttt{StatefulMigration}) and an idempotent state machine reconciler that
  replaces the Migration Manager with a Kubernetes-native control loop.

\item \textit{ShadowPod Migration Strategy:} exploits the independence of
  Kubernetes traffic routing (label selectors) from pod ownership (owner
  references) to run source and shadow pods concurrently. A shadow pod carrying
  the same application labels receives Service traffic immediately upon restore
  without violating StatefulSet identity constraints, eliminating the sequential
  stop-restore cycle and achieving zero downtime for both Deployment- and
  StatefulSet-managed workloads.

\item \textit{Identity Swap via ExchangeFence:} after ShadowPod migration
  on a StatefulSet, the shadow pod is orphaned outside the StatefulSet's
  ownership. SHADOW provides an identity swap procedure that re-checkpoints
  the shadow pod, creates a StatefulSet-owned replacement, and uses the
  ExchangeFence mechanism (unbinding, draining queues, and rebinding) to guarantee zero message loss during this handoff.
\end{enumerate}



\section{Background and Prior Work}
\label{sec:background}

\subsection{The MS2M Framework}
\label{sec:ms2m}

The Message-based Stateful Microservice Migration (MS2M)
framework~\cite{dinhtuan2022ms2m} was designed for message-driven microservices
that derive their state from processed message streams. Unlike traditional
migration approaches that treat service state as an opaque memory block, MS2M
uses a property of message-driven architectures: \textit{service state
is the result of the messages a service has processed}~\cite{laigner2025transactional}. This
makes state reconstruction possible through message replay rather than low-level
memory transfer.

%
%
%
%
%
The framework defines a five-phase migration procedure coordinated by a
Migration Manager: (1)~\textit{Checkpoint Creation}: the source container is
briefly paused using CRIU~\cite{criu} to create a process checkpoint while a
replay queue buffers incoming messages; (2)~\textit{Checkpoint Transfer}:
the archive is sent to the target host while the source continues serving;
(3)~\textit{Service Restoration}: the checkpoint is restored on the target;
(4)~\textit{Message Replay}: the restored instance processes buffered messages
from the replay queue; and (5)~\textit{Finalization}: the target switches
to the primary queue and the source is terminated.

During replay, source and target consume from separate queues, so no
message is consumed twice. Applications must suppress side effects during replay
(detectable via \texttt{START\_REPLAY}/\texttt{END\_REPLAY} control messages) or
ensure idempotent downstream operations.

The original proof-of-concept~\cite{dinhtuan2022ms2m} demonstrated a 19.92\%
reduction in service downtime compared to traditional stop-and-copy migration.

\subsection{Kubernetes Integration}
\label{sec:k8s-integration}

%
%
The integration of MS2M into Kubernetes~\cite{dinhtuan2025k8s} required
adapting the framework to the platform's declarative, desired-state resource
model. This used \textit{Forensic Container
Checkpointing} (FCC)~\cite{k8sfcc}, an experimental Kubernetes feature (since
v1.25) that extends the kubelet API to create CRIU checkpoints of running
containers without requiring direct access to the container runtime. Checkpoint
transfer was implemented by packaging the archive as an OCI-compliant container
image and pushing it to a container registry, from which the target node pulls
during pod creation. That work identified two key challenges:

\textit{Unbounded Replay Time:} When the incoming message rate approaches the
service's processing capacity, the replay queue grows indefinitely, making
migration time unpredictable. A \textit{Threshold-Based Cutoff Mechanism} was
introduced to address this: the source service is terminated after a calculated
duration $T_{\text{cutoff}} \leq \frac{T_{\text{replay\_max}} \cdot
\mu_{\text{target}}}{\lambda}$, bounding the number of messages to replay and
guaranteeing migration completion within a specified time window.

\textit{StatefulSet Constraints:} StatefulSet pods possess stable, unique network
identities that Kubernetes enforces by prohibiting two pods with the same ordinal
index from running simultaneously. This prevents the concurrent source-target
operation that MS2M relies on, forcing a sequential stop-restore cycle
(\textit{Sequential strategy}) that dominated migration time at a median of
38.5 seconds. The prior work treated this as an inherent limitation of StatefulSet
workloads.

\subsection{Kubernetes Operators}
\label{sec:operators}

A Kubernetes Operator~\cite{operatorpattern} encodes application-specific
operational knowledge as a Custom Resource Definition (CRD) and a controller
that runs a \textit{reconcile loop} to converge actual cluster state toward the
declared desired state, providing automatic retry, RBAC integration, and native
observability.

\section{System Design}
\label{sec:design}

\begin{figure*}[t]
\centering
\includegraphics[width=0.65\textwidth]{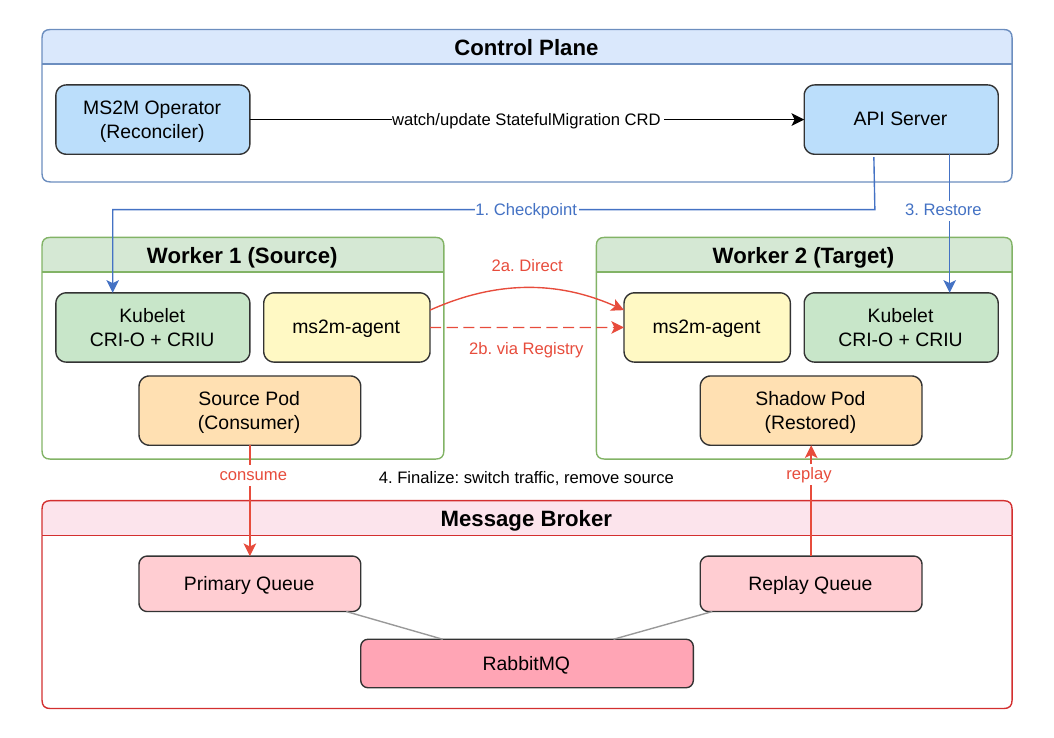}
\caption{Architecture of the SHADOW framework. The reconciler watches
StatefulMigration custom resources and orchestrates migration through the API
server: checkpoint creation via the kubelet API~(1), checkpoint
transfer between ms2m-agent instances (direct) or via registry~(2),
shadow pod creation on the target node~(3), and traffic switchover
during finalization~(4).}
\label{fig:architecture}
\end{figure*}

This section presents the architecture of SHADOW\footnote{The SHADOW implementation is publicly
available at \url{https://github.com/haidinhtuan/shadow}}  and its two optimization
strategies. Figure~\ref{fig:architecture} provides an overview of the system
components.

\subsection{The SHADOW Operator}
\label{sec:operator}

SHADOW implements the MS2M migration procedure as a Kubernetes Operator using the
\texttt{controller-runtime} framework~\cite{controllerruntime}. The operator
manages a custom resource, \texttt{StatefulMigration}, whose specification
declares the migration intent and whose status reflects the current progress.

\subsubsection{Custom Resource Definition}
\label{sec:crd}

The \texttt{StatefulMigration} CRD (API group: \texttt{migration.ms2m.io/v1alpha1})
captures the migration intent through the following key specification fields:

\begin{itemize}
\item \texttt{sourcePod}: The name of the pod to migrate.
\item \texttt{targetNode}: The destination worker node.
\item \texttt{migrationStrategy}: \texttt{Sequential} or \texttt{ShadowPod}.
  Auto-detected from the pod's owner reference chain if not specified.
\item \texttt{transferMode}: \texttt{Registry} (default) or \texttt{Direct}
  (node-to-node via ms2m-agent).
\item \texttt{messageQueueConfig}: RabbitMQ connection details, queue name, and
  exchange name for the replay mechanism.
\item \texttt{replayCutoffSeconds}: Maximum duration for the cutoff mechanism.
\end{itemize}

The resource status tracks the current \texttt{phase}, per-phase durations
(\texttt{phaseTimings}), the checkpoint identifier, and metadata cached from
the source pod (labels, container specs, owner references).

\subsubsection{State Machine Reconciler}
\label{sec:statemachine}

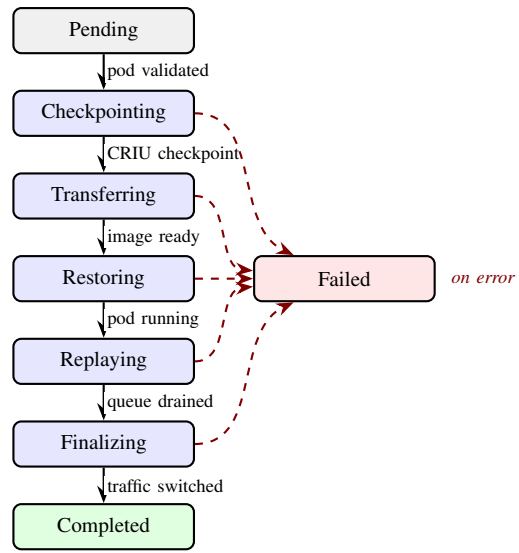
\begin{figure}[t]
\centering
\begin{tikzpicture}[
  >=Stealth,
  state/.style={draw, rounded corners=3pt, thick, minimum width=24mm,
    minimum height=6mm, font=\footnotesize},
  trans/.style={font=\scriptsize, fill=white, inner sep=1pt},
  arr/.style={->, thick},
  fail/.style={->, dashed, thick, red!50!black}
]
\node[state, fill=gray!12]  (P) at (0,0)    {Pending};
\node[state, fill=blue!10]  (C) at (0,-1.1) {Checkpointing};
\node[state, fill=blue!10]  (T) at (0,-2.2) {Transferring};
\node[state, fill=blue!10]  (R) at (0,-3.3) {Restoring};
\node[state, fill=blue!10]  (Y) at (0,-4.4) {Replaying};
\node[state, fill=blue!10]  (F) at (0,-5.5) {Finalizing};
\node[state, fill=green!12] (D) at (0,-6.6) {Completed};
\node[state, fill=red!10]   (X) at (3.2,-3.3) {Failed};

\draw[arr] (P) -- node[trans, right] {pod validated} (C);
\draw[arr] (C) -- node[trans, right] {CRIU checkpoint} (T);
\draw[arr] (T) -- node[trans, right] {image ready} (R);
\draw[arr] (R) -- node[trans, right] {pod running} (Y);
\draw[arr] (Y) -- node[trans, right] {queue drained} (F);
\draw[arr] (F) -- node[trans, right] {traffic switched} (D);

\draw[fail] (C.east) to[out=0,in=155] (X);
\draw[fail] (T.east) to[out=0,in=175] (X);
\draw[fail] (R.east) -- (X);
\draw[fail] (Y.east) to[out=0,in=185] (X);
\draw[fail] (F.east) to[out=0,in=205] (X);
\node[font=\scriptsize\itshape, text=red!50!black, anchor=west]
  at (X.east) {~on error};
\end{tikzpicture}
\caption{State machine of the StatefulMigration reconciler. The reconcile loop
dispatches to the handler for the current phase, advancing on success or
transitioning to Failed on error. Each handler is idempotent for safe retry.}
\label{fig:statemachine}
\end{figure}

The reconciler implements the MS2M five-phase procedure as an idempotent state
machine with seven phases, illustrated in Figure~\ref{fig:statemachine}:
\texttt{Pending}, \texttt{Checkpointing},
\texttt{Transferring}, \texttt{Restoring}, \texttt{Replaying},
\texttt{Finalizing}, and \texttt{Completed} (or \texttt{Failed}). Each
reconciliation invocation dispatches to the handler for the current phase, which
either advances to the next phase or re-queues for a later attempt.

Phase handlers are \textit{idempotent}: re-entering a phase after a temporary
failure does not corrupt migration state. The reconciler implements
\textit{phase chaining} (executing synchronously-completed phases within a
single API call) and \textit{exponential polling backoff} for long-running
phases, reducing API server load. Latencies for each phase are
recorded in \texttt{status.phaseTimings} for built-in instrumentation.

When \texttt{migrationStrategy} is not specified, the operator auto-detects it
from the pod's owner reference chain (StatefulSet defaults to Sequential, Deployment
defaults to ShadowPod), but this can be overridden explicitly.

\subsection{ShadowPod Migration Strategy}
\label{sec:shadowpod}

The ShadowPod strategy removes the sequential restore bottleneck by allowing
source and target pods to operate concurrently during the replay phase.
Kubernetes Services route traffic based on \textit{label selectors}, while
StatefulSet identity constraints are enforced through \textit{owner references}.
A shadow pod that carries the same application
labels but is not owned by the StatefulSet can coexist with the original pod
and receive Service traffic without violating identity constraints.

\subsubsection{Core Procedure}
\label{sec:shadowpod-core}

\begin{figure}[t]
\centering
\includegraphics[width=\columnwidth]{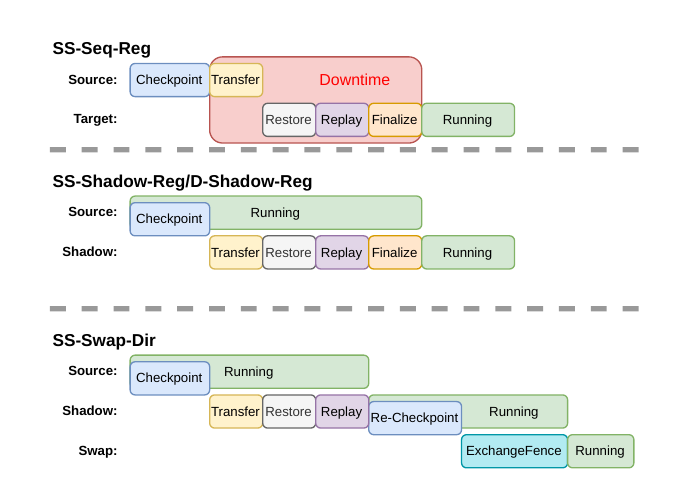}
\caption{Timeline comparison of the migration strategies.
SS-Seq-Reg: the service is unavailable throughout the Restoring phase.
SS-Shadow-Reg/D-Shadow-Reg: the source continues serving while the shadow pod restores
and replays. SS-Swap-Dir: after replay, the shadow pod continues serving during
the identity swap (re-checkpoint, ExchangeFence), and is replaced only once
the StatefulSet-owned replacement is ready.}
\label{fig:timeline}
\end{figure}

Figure~\ref{fig:timeline} illustrates the key differences between the three
StatefulSet strategies. In Sequential, the source pod must be terminated before
the target can be created, causing a service downtime. In ShadowPod, the shadow pod
runs alongside the source. In ShadowPod+Swap, an additional identity swap phase
replaces the shadow pod with a StatefulSet-owned replacement.

The ShadowPod strategy modifies the Restoring and Finalizing phases of the
migration procedure. During the \textit{Restoring} phase, instead of scaling
down a StatefulSet and recreating its pod, the operator creates a new pod
named \texttt{<source>-shadow} on the target node. This shadow pod uses the
checkpoint container image with \texttt{imagePullPolicy: Never} (since the
image is already present in the target node's local image store). The shadow
pod carries the same application labels as the source pod, so it can
receive traffic via the Kubernetes Service immediately upon becoming ready.
Both the source pod and the shadow pod run concurrently during this phase
and throughout the subsequent replay phase.

Because the shadow pod is restored from a CRIU checkpoint (which includes the
running HTTP server thread), its health endpoint is available immediately after
restore. Kubernetes marks the shadow pod as Ready and adds it to the Service's
endpoint list, so both pods serve traffic during replay. The source provides
fully up-to-date state while the shadow pod's state converges as replay
progresses.

\subsubsection{Finalization by Workload Type}
\label{sec:shadowpod-deployment}
\label{sec:shadowpod-statefulset}

For Deployment-managed pods, finalization sends \texttt{END\_REPLAY} to the
shadow pod, deletes the source pod, patches the Deployment's
\texttt{nodeAffinity} to the target node, and cleans up the replay queue.
For StatefulSet-managed pods, the StatefulSet is scaled down by one replica
instead of deleting the source directly (which would cause the controller to
recreate it). The StatefulSet controller removes the highest-ordinal pod
(the source), while the shadow pod continues serving traffic.

\subsubsection{Trade-offs}
\label{sec:shadowpod-tradeoffs}

For StatefulSets, the ShadowPod strategy results in the workload being served
by a standalone shadow pod outside StatefulSet ownership. StatefulSet guarantees
(ordered scaling, PVC attachments) do not apply. SHADOW therefore targets
in-memory stateful services whose state is derived from message processing, not
persistent disk storage. For Deployment-managed workloads, no such trade-off
exists.

\subsection{Identity Swap via ExchangeFence}
\label{sec:identity-swap}

The ShadowPod strategy for StatefulSets leaves the workload served by an orphaned shadow pod outside the StatefulSet's ownership. To restore full StatefulSet management, SHADOW provides an identity swap procedure during the Finalizing phase:

\begin{enumerate}
\item The shadow pod is re-checkpointed on the target node using a local CRIU dump.
\item The StatefulSet is scaled down by one replica, removing the original source pod.
\item A replacement pod with the original name (e.g., \texttt{consumer-0}) is created from the re-checkpoint image. The StatefulSet controller adopts this pod.
\item The shadow pod is terminated.
\end{enumerate}

During step~4, traffic must be handed off from the shadow pod to the replacement pod without message loss. A naive cutoff (terminating the shadow and starting the replacement) risks duplicate processing or lost messages if the handoff is not atomic. SHADOW addresses this with the \textit{ExchangeFence} mechanism:

\begin{enumerate}
\item A temporary buffer queue is bound to the message exchange.
\item The exchange is unbound from both the primary queue and the replay queue, fencing new messages into the buffer.
\item Both the primary queue and the replay queue are drained to zero.
\item The shadow pod is terminated, the primary queue is rebound to the exchange, and the buffer queue is drained into the replacement pod.
\end{enumerate}

The ExchangeFence provides a consistent cut: no messages are lost because the buffer queue captures all messages during the fence window, and no messages are duplicated because both source queues are fully drained before the handoff completes. This procedure adds latencies to the Finalizing phase (depending on message rate and queue depths) but restores full StatefulSet ownership of the migrated workload.

\subsection{Direct Node-to-Node Transfer}
\label{sec:direct-transfer}

The second optimization targets the container registry round-trip during
checkpoint transfer. The \texttt{ms2m-agent} DaemonSet enables direct
worker-to-worker transfer that bypasses the registry entirely: the source node
streams the checkpoint archive via HTTP to the target agent, which constructs the
OCI image locally and loads it into the container runtime's image store.
Three of the four evaluation configurations (SS-Seq-Reg, SS-Shadow-Reg, D-Shadow-Reg)
use registry-based transfer; SS-Swap-Dir uses agent-assisted direct transfer for
both the initial checkpoint and the identity-swap re-checkpoint.

\subsubsection{The ms2m-agent DaemonSet}
\label{sec:agent}

The \texttt{ms2m-agent} is deployed as a Kubernetes DaemonSet on all worker nodes.
The agent exposes an HTTP endpoint (port 9443) that accepts checkpoint archives
and performs local image construction. The agent's pod specification includes
\texttt{hostPath} volume mounts for \texttt{/var/lib/kubelet/checkpoints}
(read-only, for accessing FCC checkpoint archives) and \texttt{/var/lib/ms2m}
(read-write, for temporary image construction).

\subsubsection{Transfer Procedure}
\label{sec:transfer-procedure}

When \texttt{transferMode: Direct} is specified, the Transferring phase operates
as follows:

\begin{enumerate}
\item The operator creates a transfer Job scheduled on the source node.
\item The Job reads the checkpoint archive from the kubelet's checkpoint
  directory.
\item The Job streams the archive as a multipart HTTP POST to the target
  node's ms2m-agent at
  \nolinkurl{http://ms2m-agent.ms2m-system.svc:9443/checkpoint}.
\item The target agent writes the archive to disk, constructs an
  OCI-compliant container image (a single uncompressed layer with CRI-O
  checkpoint annotations), and loads it into the local image store using
  \texttt{skopeo copy}.
\item The agent responds with the local image reference, which the operator
  records in the migration status.
\end{enumerate}

The OCI image uses an uncompressed layer, as the CPU cost of compression outweighs
bandwidth savings on cluster-local networks. This direct path eliminates the
registry push/pull round-trip. The default \texttt{Registry} transfer mode
from~\cite{dinhtuan2025k8s} remains available as a fallback.

\subsubsection{CRIU Hostname Resolution}
\label{sec:hostname}

After CRIU restore, \texttt{gethostname()} returns the source pod's name (from
the UTS (UNIX Time-Sharing) namespace snapshot), causing the control message protocol to listen on
the wrong queue. This is resolved by reading \texttt{/etc/hostname} (bind-mounted
by the runtime, reflecting the actual pod name) instead.

\section{Evaluation}
\label{sec:evaluation}

This section presents the experimental evaluation of SHADOW across four
migration configurations, comparing the baseline Sequential approach with the
ShadowPod strategy on both StatefulSet and Deployment workloads.

\subsection{Experimental Setup}
\label{sec:setup}

\subsubsection{Infrastructure}

The evaluation was conducted on a bare-metal Kubernetes cluster provisioned on
a European cloud provider, consisting of three dedicated servers:

\begin{itemize}
\item \textit{Control plane:} 1 node running the Kubernetes API server,
  scheduler, controller manager, and the MS2M operator.
\item \textit{Workers:} 2 nodes serving as alternating source and target for
  round-trip migrations.
\end{itemize}

Each server is equipped with 4 dedicated vCPUs, 8\,GB RAM, and 80\,GB SSD
storage. All nodes run Ubuntu 22.04 with Kubernetes v1.32, CRI-O as the
container runtime, and CRIU v4.0 compiled from source for checkpoint/restore
operations. An in-cluster container registry (deployed in the \texttt{registry}
namespace) is used for OCI checkpoint image transfer. RabbitMQ 3.13 is deployed
as a single-instance StatefulSet for message brokering.

\subsubsection{Workload}

The evaluation workload consists of two microservices:

\begin{itemize}
\item \textit{Producer:} A Go Deployment that publishes messages to a RabbitMQ
  fanout exchange at a configurable rate. Each message contains a sequence
  number and timestamp.
\item \textit{Consumer:} A single-replica workload (StatefulSet or Deployment,
  depending on configuration) implemented in Python. The consumer maintains an
  in-memory counter of processed messages and the last-seen sequence number as
  its stateful context. It exposes an HTTP health endpoint on port 8080 that
  reports the current processing state, serving both as a readiness probe and
  a downtime measurement target.
\end{itemize}

The consumer implements the MS2M control message protocol
(Section~\ref{sec:ms2m}) for queue switching during replay.

\subsubsection{Configurations}

Four migration configurations are evaluated to assess the ShadowPod strategy,
identity swap, and cross-workload applicability:

\begin{enumerate}
\item \textit{SS-Seq-Reg} (statefulset-sequential-register): Baseline
  from~\cite{dinhtuan2025k8s}. StatefulSet consumer, Sequential strategy,
  registry transfer.
\item \textit{SS-Shadow-Reg} (statefulset-shadowpod-register): StatefulSet consumer,
  ShadowPod strategy, registry transfer. Isolates the ShadowPod effect.
\item \textit{SS-Swap-Dir} (statefulset-shadowpod-swap-direct): StatefulSet consumer,
  ShadowPod strategy with ExchangeFence identity swap
  (Section~\ref{sec:identity-swap}), agent-assisted direct transfer
  (Section~\ref{sec:direct-transfer}). Both the initial checkpoint and the
  identity-swap re-checkpoint are loaded directly into the target node's
  container storage, bypassing the registry entirely.
\item \textit{D-Shadow-Reg} (deployment-shadowpod-registry): Deployment consumer, ShadowPod strategy, registry transfer.
\end{enumerate}

The comparison between SS-Seq-Reg and SS-Shadow-Reg isolates the effect of the
ShadowPod strategy on the same workload type (StatefulSet). The comparison
between SS-Shadow-Reg and SS-Swap-Dir reflects two simultaneous changes: the identity
swap procedure and the switch from registry to agent-assisted direct transfer;
the lower Transferring time of SS-Swap-Dir (~0.19\,s vs.~5.6\,s) is attributable
to the direct transfer mode, not to the identity swap itself. The comparison
between SS-Shadow-Reg and D-Shadow-Reg reveals the effect of workload type (StatefulSet
vs.\ Deployment) under the same ShadowPod strategy and registry transfer mode.

\subsubsection{Parameters}

Each configuration is evaluated at seven message rates: 10, 20, 40, 60, 80,
100, and 120 messages per second. Each rate--configuration combination is
repeated 10 times, yielding $4 \times 7 \times 10 = 280$ total migration runs.
The replay cutoff is set to 120 seconds. Between runs, message queues are purged
to prevent accumulation, checkpoint images are cleaned from worker nodes to
prevent disk exhaustion, and the consumer pod is verified to be ready and
processing messages before initiating the next migration.

\subsubsection{Downtime Measurement}
\label{sec:downtime-method}

Service downtime is measured using an external HTTP probe pod that sends
requests to the consumer's Service endpoint at 10\,ms intervals. Downtime is
the \textit{longest contiguous streak} of failed probes during the migration
window, with a 3\,s gap threshold to avoid merging unrelated failures. Total migration time is the elapsed time from
\texttt{StatefulMigration} resource creation to the Completed phase.

\subsection{Results}
\label{sec:results}

\subsubsection{Total Migration Time}

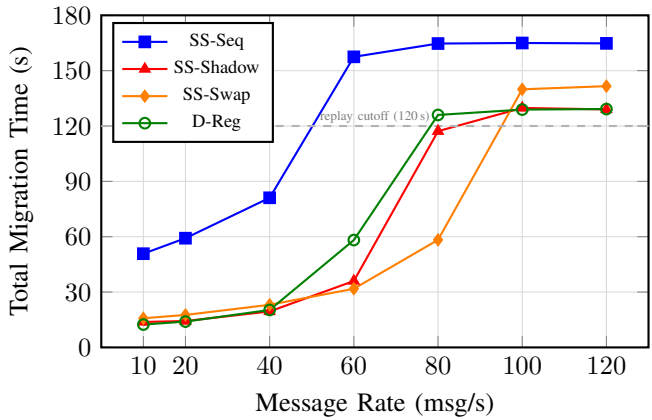
\begin{figure}[t]
\centering
\begin{tikzpicture}
\begin{axis}[
  width=\columnwidth,
  height=6cm,
  ylabel={Total Migration Time (s)},
  xlabel={Message Rate (msg/s)},
  xmin=0, xmax=130,
  ymin=0, ymax=180,
  xtick={10,20,40,60,80,100,120},
  ytick={0,30,60,90,120,150,180},
  grid=major,
  grid style={gray!30},
  legend style={at={(0.02,0.98)}, anchor=north west, font=\scriptsize,
    legend columns=1},
  mark size=2pt,
  thick,
]
\addplot[mark=square*, blue] coordinates {
  (10,50.8) (20,59.2) (40,81.1) (60,157.5) (80,164.7) (100,165.0) (120,164.8)};
\addplot[mark=triangle*, red] coordinates {
  (10,13.8) (20,14.3) (40,19.5) (60,36.0) (80,117.2) (100,129.8) (120,129.0)};
\addplot[mark=diamond*, orange] coordinates {
  (10,15.8) (20,17.6) (40,23.1) (60,31.8) (80,58.3) (100,139.9) (120,141.6)};
\addplot[mark=o, green!50!black] coordinates {
  (10,12.4) (20,14.0) (40,20.3) (60,58.2) (80,126.0) (100,128.9) (120,129.2)};
\addplot[dashed, gray!60, forget plot] coordinates {(0,120) (130,120)};
\node[font=\tiny, gray] at (axis cs:65,125) {replay cutoff (120\,s)};
\legend{SS-Seq-Reg, SS-Shadow-Reg, SS-Swap-Dir, D-Shadow-Reg}
\end{axis}
\end{tikzpicture}
\caption{Total migration time (median, $n=10$) across message rates.
At low rates ($\leq$\,40\,msg/s), ShadowPod reduces time by 73--76\%. At
$\geq$\,100\,msg/s, where the 120\,s replay cutoff dominates, the reduction
narrows to 15--22\%. The dashed line marks the replay cutoff boundary.}
\label{fig:total-time}
\end{figure}

Table~\ref{tab:total-time} presents the median total migration time
($n=10$) for each configuration across all seven message rates.
Figure~\ref{fig:total-time} visualizes the trend.

%
At low rates (10\,msg/s), SS-Shadow-Reg and D-Shadow-Reg complete in
12.4--13.8\,s, a 73--76\% reduction from the Sequential baseline of 50.8\,s,
primarily by eliminating the 38.5\,s median restore phase. At
$\geq$\,100\,msg/s, where the replay cutoff dominates all configurations, the
ShadowPod advantage narrows to 22\%. At the intermediate rate of 60\,msg/s, SS-Shadow-Reg
completes in 36.0\,s vs.\ 157.5\,s for Sequential (77\% reduction).

The SS-Swap-Dir configuration shows similar total times to SS-Shadow-Reg at low rates
(15.8\,s vs.\ 13.8\,s at 10\,msg/s), with the additional 2\,s attributable to
the identity swap finalization. At high rates, SS-Swap-Dir is 12.6\,s slower than
SS-Shadow-Reg (141.6\,s vs.\ 129.0\,s at 120\,msg/s) due to the 17.5\,s finalize
phase, but still 14\% faster than Sequential.

\begin{table*}[t]
\centering
\caption{Total Migration Time, median (seconds), $n=10$. Values
in parentheses indicate the interquartile range (Q1--Q3).}
\label{tab:total-time}
\begin{tabular}{lcccc}
\toprule
\textbf{Rate} & \textbf{\shortstack{SS-Seq\\-Reg}} & \textbf{\shortstack{SS-Shadow\\-Reg}} & \textbf{\shortstack{SS-Swap\\-Dir}} & \textbf{\shortstack{D-Shadow\\-Reg}} \\
\midrule
10  & 50.8\,\scriptsize(50.3--80.1) & 13.8\,\scriptsize(12.3--13.8) & 15.8\,\scriptsize(14.9--17.4) & 12.4\,\scriptsize(12.2--13.3) \\
20  & 59.2\,\scriptsize(58.2--60.0) & 14.3\,\scriptsize(13.3--44.3) & 17.6\,\scriptsize(16.7--20.4) & 14.0\,\scriptsize(13.0--16.3) \\
40  & 81.1\,\scriptsize(80.2--105.9) & 19.5\,\scriptsize(18.7--84.8) & 23.1\,\scriptsize(21.9--28.4) & 20.3\,\scriptsize(18.9--29.4) \\
60  & 157.5\,\scriptsize(130.1--164.5) & 36.0\,\scriptsize(29.1--100.5) & 31.8\,\scriptsize(30.0--32.8) & 58.2\,\scriptsize(27.9--108.3) \\
80  & 164.7\,\scriptsize(163.9--165.2) & 117.2\,\scriptsize(74.7--129.3) & 58.3\,\scriptsize(52.5--67.5) & 126.0\,\scriptsize(103.4--129.7) \\
100 & 165.0\,\scriptsize(164.5--165.9) & 129.8\,\scriptsize(128.9--130.9) & 139.9\,\scriptsize(138.9--140.2) & 128.9\,\scriptsize(128.4--130.0) \\
120 & 164.8\,\scriptsize(164.3--165.5) & 129.0\,\scriptsize(128.5--130.1) & 141.6\,\scriptsize(141.1--142.0) & 129.2\,\scriptsize(128.9--131.4) \\
\bottomrule
\end{tabular}
\end{table*}

\subsubsection{Phase-by-Phase Breakdown}

\begin{figure}[t]
\centering
\includegraphics[width=\columnwidth]{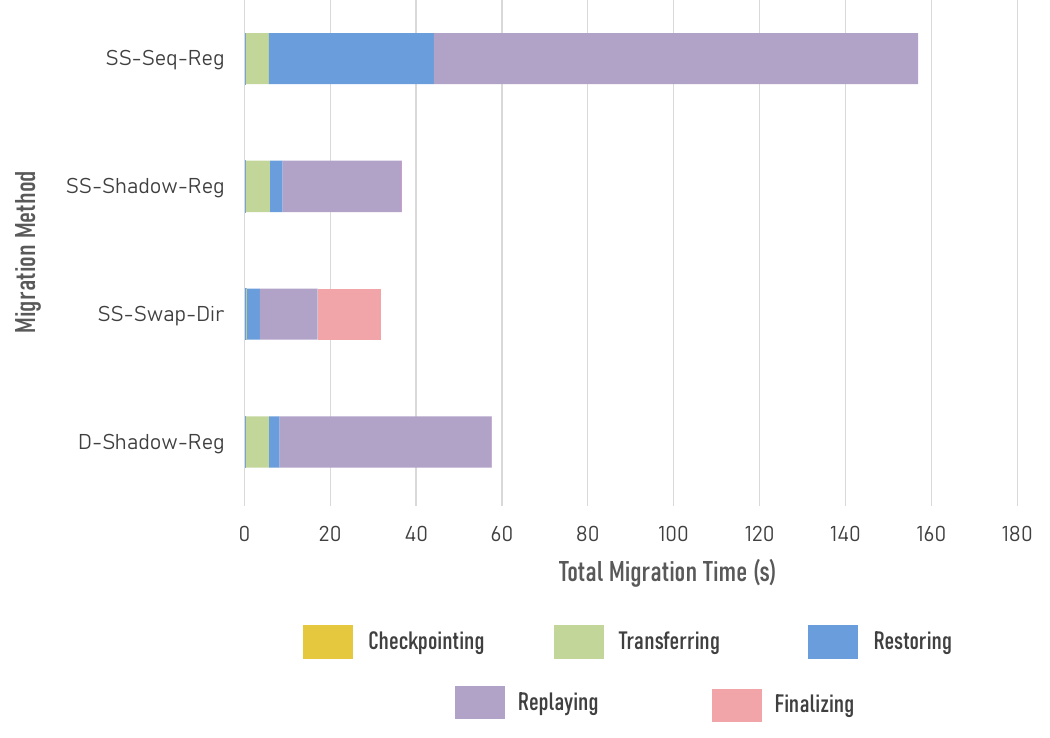}
\caption{Phase duration breakdown at 60\,msg/s (medians, $n=10$). The
SS-Seq-Reg configuration has a long 38.41\,s restore phase. ShadowPod
reduces restore to 2.50--3.02\,s, shifting the bottleneck to the replay phase.}
\label{fig:phase-breakdown}
\end{figure}

Tables~\ref{tab:phases-10} and~\ref{tab:phases-60} present the per-phase
durations at 10 and 60\,msg/s respectively. \textbf{Checkpointing} is consistent at a median of 0.34\,s across all
configurations. \textbf{Transferring} takes 5.26--5.64\,s for registry-based
configurations (SS-Seq-Reg, SS-Shadow-Reg, D-Shadow-Reg), as the phase is I/O-bound by the OCI
image push and pull; SS-Swap-Dir reduces this to 0.19--0.20\,s via agent-assisted
local load, bypassing the registry entirely. The \textbf{Restoring} phase shows
the primary ShadowPod benefit: Sequential requires a median of 38.41--38.49\,s
(StatefulSet identity constraint), while ShadowPod requires only 2.28--3.23\,s, a
92\% reduction. The \textbf{Replaying} phase scales with message rate: at
60\,msg/s, SS-Shadow-Reg replays in 27.68\,s vs.\ 112.80\,s for Sequential, because
fewer messages accumulate during the shorter restore window. At
$\geq$\,100\,msg/s, all configurations hit the 120\,s cutoff. At 80\,msg/s,
SS-Swap-Dir's median replay (37.9\,s) remains well below the cutoff owing to its
shorter restore window and lower initial queue depth.
\textbf{Finalizing} is near-instantaneous ($<$\,0.1\,s) for SS-Seq-Reg, SS-Shadow-Reg,
and D-Shadow-Reg; SS-Swap-Dir incurs a substantially longer Finalizing phase (8.64\,s at
10\,msg/s, 14.73\,s at 60\,msg/s) due to the identity swap procedure:
re-checkpointing, pod replacement, and ExchangeFence queue draining.
Figure~\ref{fig:phase-breakdown} visualizes the breakdown at 60\,msg/s.

\begin{table}[t]
\centering
\caption{Phase Durations (median seconds, $n=10$) at 10\,msg/s}
\label{tab:phases-10}
\begin{tabular}{@{}lcccc@{}}
\toprule
\textbf{Phase} & \textbf{\shortstack{SS-Seq\\-Reg}} & \textbf{\shortstack{SS-Shadow\\-Reg}} & \textbf{\shortstack{SS-Swap\\-Dir}} & \textbf{\shortstack{D-Shadow\\-Reg}} \\
\midrule
Checkpointing  & 0.33  & 0.33  & 0.37 & 0.34 \\
Transferring   & 5.30  & 5.60  & 0.20 & 5.26 \\
Restoring      & 38.49 & 3.23  & 2.61 & 2.28 \\
Replaying      & 6.53  & 4.08  & 3.78 & 4.30 \\
Finalizing     & 0.01  & 0.02  & 8.64 & 0.03 \\
\midrule
\textbf{Total} & 50.8  & 13.8  & 15.8 & 12.4 \\
\bottomrule
\end{tabular}
\end{table}

\begin{table}[t]
\centering
\caption{Phase Durations (median seconds, $n=10$) at 60\,msg/s}
\label{tab:phases-60}
\begin{tabular}{@{}lcccc@{}}
\toprule
\textbf{Phase} & \textbf{\shortstack{SS-Seq\\-Reg}} & \textbf{\shortstack{SS-Shadow\\-Reg}} & \textbf{\shortstack{SS-Swap\\-Dir}} & \textbf{\shortstack{D-Shadow\\-Reg}} \\
\midrule
Checkpointing  & 0.34  & 0.34  & 0.39 & 0.33 \\
Transferring   & 5.35  & 5.64  & 0.19 & 5.38 \\
Restoring      & 38.41 & 2.89  & 3.02 & 2.50 \\
Replaying      & 112.80 & 27.68 & 13.44 & 49.33 \\
Finalizing     & 0.00  & 0.02  & 14.73 & 0.04 \\
\midrule
\textbf{Total} & 157.5 & 36.0  & 31.8 & 58.2 \\
\bottomrule
\end{tabular}
\end{table}

\subsubsection{Service Downtime}

Table~\ref{tab:downtime} reports the measured service downtime (median, $n=10$)
using the HTTP probe methodology described in
Section~\ref{sec:downtime-method}.

The Sequential baseline exhibits a consistent median of approximately 31\,s of downtime across
all rates (rate-independent, determined by the StatefulSet scale-down/up cycle).
All three ShadowPod configurations achieve zero measured downtime
across all rates and all 210 ShadowPod runs (70 per configuration).


\begin{table}[t]
\centering
\caption{Service Downtime (median) by Configuration and Message Rate ($n=10$).}
\label{tab:downtime}
\begin{tabular}{@{}lcccc@{}}
\toprule
\textbf{Rate} & \textbf{\shortstack{SS-Seq\\-Reg}} & \textbf{\shortstack{SS-Shadow\\-Reg}} & \textbf{\shortstack{SS-Swap\\-Dir}} & \textbf{\shortstack{D-Shadow\\-Reg}} \\
\midrule
10  & 31.2\,s & 0\,ms & 0\,ms & 0\,ms \\
20  & 31.2\,s & 0\,ms & 0\,ms & 0\,ms \\
40  & 31.2\,s & 0\,ms & 0\,ms & 0\,ms \\
60  & 31.2\,s & 0\,ms & 0\,ms & 0\,ms \\
80  & 31.1\,s & 0\,ms & 0\,ms & 0\,ms \\
100 & 31.1\,s & 0\,ms & 0\,ms & 0\,ms \\
120 & 30.9\,s & 0\,ms & 0\,ms & 0\,ms \\
\bottomrule
\end{tabular}
\end{table}

\subsubsection{Message Loss}

Across all 280 migration runs, no messages were lost during migration. The
message replay mechanism successfully synchronized state between source and
target in every completed run across all four configurations and all seven
message rates, confirming the zero-loss guarantee of the MS2M framework.

\subsection{Discussion}
\label{sec:discussion}

\subsubsection{Interpretation}


The comparison between configurations~1 and~2 (SS-Seq-Reg vs.\ SS-Shadow-Reg)
isolates the effect of the ShadowPod strategy on the same workload type (StatefulSet). The 92\% reduction
in restore duration confirms that the StatefulSet identity constraint, which
prior work treated as an inherent limitation, can be circumvented by
decoupling traffic routing (label selectors) from pod ownership (owner
references). Kubernetes Services and StatefulSet identity operate on
independent mechanisms, and this separation is what makes zero-downtime
migration possible without requiring a workload type change.

The comparison between configurations~2 and~3 (SS-Shadow-Reg vs.\ SS-Swap-Dir)
reflects two simultaneous changes: the identity swap procedure and the switch to
agent-assisted direct transfer. SS-Swap-Dir's transfer phase is approximately 30x faster than
SS-Shadow-Reg (0.19\,s vs.\ 5.64\,s at 60\,msg/s) because agent-assisted
direct transfer bypasses the registry entirely, but its finalize phase is approximately 750x longer (14.73\,s vs.\ 0.02\,s) due to
the re-checkpoint, pod replacement, and ExchangeFence procedure. The net effect
is a comparable total migration time (31.8\,s vs.\ 36.0\,s at 60\,msg/s
for SS-Swap-Dir vs.\ SS-Shadow-Reg), with the benefit of restoring full StatefulSet
ownership of the migrated workload.

The comparison between configurations~2 and~4 (SS-Shadow-Reg vs.\ D-Shadow-Reg) shows
that the ShadowPod strategy applies to both StatefulSet and Deployment
workloads, with comparable restore times (2.89\,s vs.\ 2.50\,s at 60\,msg/s)
and zero downtime in both cases. The replay duration differs more noticeably
(27.7\,s vs.\ 49.3\,s at 60\,msg/s). Both shadow pods start draining the replay
queue immediately after restore, but D-Shadow-Reg drains more slowly at
60\,msg/s, suggesting a lower effective message processing rate for the
Deployment-based consumer under that load. At lower rates ($\leq$\,40\,msg/s),
replay times are comparable, confirming that the divergence
is rate-dependent rather than a fixed overhead.

At high message rates ($\geq$\,100\,msg/s), the replay cutoff dominates total
migration time across all configurations. Even at 80\,msg/s, the three
registry-based configurations approach the cutoff (medians 108--120\,s replay).
This reveals a limitation of
the MS2M replay mechanism: when $\mu_{\text{target}} < \lambda$, the replay
queue grows faster than it can be drained, and the cutoff fires with messages
still pending. The ShadowPod strategy does \textit{reduce} the replay
queue length compared to Sequential (because the source continues processing
during the shorter restore window), but cannot eliminate the rate-dependent
bottleneck. Techniques such as batched message processing, increased consumer
parallelism, or CRIU pre-dump for incremental checkpointing would address this
bottleneck.

Zero message loss across all 280 runs confirms the correctness of the
concurrent source-shadow operation: the fanout exchange ensures that both
pods receive identical message streams on separate queues, and the control
message protocol guarantees an orderly handoff.

\subsubsection{Comparison with Prior Work}

%
The prior evaluation~\cite{dinhtuan2025k8s} used GCE e2-medium VMs with a
Java consumer. Despite different infrastructure, checkpoint (median 0.34\,s
vs.\ 0.4\,s), transfer (median 5.4\,s vs.\ 6\,s), and
Sequential restore (median 38.5\,s vs.\ 39\,s) durations are
comparable, confirming these phases are I/O-bound and the restore bottleneck is
intrinsic to StatefulSet identity constraints.

\subsubsection{Trade-offs and Operator Overhead}

The ShadowPod strategy for StatefulSets leaves the workload on a standalone
shadow pod; Sequential remains available when full StatefulSet guarantees are
required. Direct transfer (SS-Swap-Dir) reduces transfer time from~5.6\,s
to~0.19\,s at the cost of deploying the ms2m-agent DaemonSet. Operator
reconcile loop overhead is negligible relative to phase durations.
\section{Related Work}
\label{sec:related}

Container live migration techniques are surveyed by Soussi
et al.~\cite{soussi2024survey}. Pre-copy transfers dirty pages iteratively,
trading total time for lower downtime~\cite{lu2023mbdpc}; post-copy resumes
immediately but degrades performance during page
faults~\cite{guitart2024postcopy}. SHADOW avoids this trade-off by using a point-in-time CRIU snapshot
combined with message replay, eliminating the need for iterative dirty-page
synchronization.
FCC~\cite{k8sfcc} provides the kubelet API for CRIU checkpointing used by SHADOW.
KubeSPT~\cite{kubespt2025} operates directly on the container runtime (Docker)
rather than the kubelet FCC API. It addresses
stateful pod migration through iterative pre-copy checkpointing (T-Checkpointer),
TCP connection preservation via network namespace freezing (T-Proxy), and
hot-data-first restoration with lazy-loading of remaining pages (T-Restorer),
reporting 86--93\% downtime reduction for memory-intensive workloads such as Redis.
KubeSPT focuses on raw memory and network state preservation and does not address
application-level consistency for message-driven services. SHADOW instead
reconstructs state through message replay, achieving true zero-downtime by keeping
the source pod serving throughout, whereas KubeSPT still incurs downtime during the
final checkpoint and restore window.

Other CRIU-based approaches include Guitart's~\cite{guitart2024postcopy}
diskless, iterative (pre-copy and post-copy), and connection-persistent live
migrations of runC containers for HPC workloads; Voyager's~\cite{ums} complete
state transfer combining CRIU memory migration with filesystem data federation
and lazy replication; Ma et al.'s~\cite{ma2019live} filesystem-layer transfer
for edge (56--80\% time reduction); and Calagna et al.'s~\cite{calagna2024coat}
COAT (TCP connection preservation via OvS overlay) and PAM (migration KPI
prediction model) for Podman-based edge workloads. None support concurrent
source-target operation, which is what makes SHADOW's zero-downtime guarantee
possible.
The Operator pattern~\cite{operatorpattern} has been widely adopted for managing
complex application lifecycles in Kubernetes, including database operators
(e.g., for PostgreSQL, MySQL) and stateful middleware. Both
KubeSPT~\cite{kubespt2025} and SHADOW use Custom Resource Definitions to drive
migration as a Kubernetes-native workflow. SHADOW extends this pattern with an
idempotent state machine reconciler that supports multiple migration strategies
(Sequential, ShadowPod) and transfer modes (Registry, Direct), supporting
declarative migration orchestration within Kubernetes' desired-state model.

\section{Conclusion}
\label{sec:conclusion}
SHADOW shows that Kubernetes' StatefulSet identity constraint, previously
treated as an unavoidable source of migration downtime, can be circumvented at the
application level by separating traffic routing (label-based) from pod ownership
(controller-based). The ShadowPod strategy uses this separation for
concurrent source-target operation,
eliminating service downtime entirely while preserving zero message loss. This
result holds across both Deployment-managed and StatefulSet-managed workloads,
confirming that the approach generalizes beyond a single workload controller.

For StatefulSet workloads that require full controller ownership after migration,
the identity swap procedure with ExchangeFence re-checkpoints the shadow pod and
creates a StatefulSet-adopted replacement, adding overheads but
restoring ordered scaling and crash recovery guarantees.

Beyond the performance improvements (92\% restore reduction, up to 77\% total time
reduction), encoding migration as a Kubernetes-native control loop provides
automatic failure recovery, declarative lifecycle management, and built-in
observability, properties difficult to achieve with external scripts, confirmed
across 280 migration runs with zero failures.

The remaining performance bottleneck is the replay phase at high message rates,
where the cutoff mechanism bounds migration time but leaves the shadow pod with
partially-synchronized state. Future work can address this through CRIU
pre-dump support for incremental checkpointing (reducing both checkpoint size
and transfer time), batched message replay to increase effective consumer
throughput, and extension of migration support to multi-container pods.


\balance
\bibliographystyle{IEEEtran}
\bibliography{references}

\end{document}